\pdfoutput=1
\documentclass[sigconf, screen=true]{acmart}
\usepackage[T1]{fontenc}
\usepackage[utf8]{inputenc}

\copyrightyear{2021}
\acmYear{2021}
\setcopyright{rightsretained}
\acmConference[ICMI '21]{Proceedings of the 2021 International Conference on Multimodal Interaction}{October 18--22, 2021}{Montréal, QC, Canada}
\acmBooktitle{Proceedings of the 2021 International Conference on Multimodal Interaction (ICMI '21), October 18--22, 2021, Montréal, QC, Canada}
\acmDOI{10.1145/3462244.3479957}
\acmISBN{978-1-4503-8481-0/21/10}

\usepackage{booktabs}
\usepackage{caption}
\usepackage{subcaption}
\usepackage{graphicx}
\usepackage{pifont}
\newcommand{\cmark}{\ding{51}}
\newcommand{\xmark}{\ding{55}}

\usepackage{enumitem} 
\setlist{nolistsep} 

\begin{document}

\title{HEMVIP: Human Evaluation of Multiple Videos in Parallel}

\author{Patrik Jonell}
\email{pjjonell@kth.se}
\affiliation{%
  \institution{Speech, Music and Hearing,\\{}KTH Royal Institute of Technology}
  \city{Stockholm}
  \country{Sweden}
}

\author{Youngwoo Yoon}
\email{youngwoo@etri.re.kr}
\affiliation{%
  \institution{ETRI \& KAIST}
  \city{Daejeon}
  \country{Republic of Korea}
}

\author{Pieter Wolfert}
\email{pieter.wolfert@ugent.be}
\affiliation{%
 \institution{IDLab, Ghent University -- imec}
 \city{Ghent}
 \country{Belgium}}
 
 \author{Taras Kucherenko}
\email{tarask@kth.se}
\affiliation{%
  \institution{Robotics, Perception and Learning, KTH Royal Institute of Technology}
  \city{Stockholm}
  \country{Sweden}
}

\author{Gustav Eje Henter}
\email{ghe@kth.se}
\affiliation{%
  \institution{Speech, Music and Hearing,\\{}KTH Royal Institute of Technology}
  \city{Stockholm}
  \country{Sweden}
}


\begin{abstract}
In many research areas, for example motion and gesture generation, objective measures alone do not provide an accurate impression of key stimulus traits such as perceived quality or appropriateness. The gold standard is instead to evaluate these aspects through user studies, especially subjective evaluations of video stimuli. Common evaluation paradigms either present individual stimuli to be scored on Likert-type scales, or ask users to compare and rate videos in a pairwise fashion. However, the time and resources required for such evaluations scale poorly as the number of conditions to be compared increases. Building on standards used for evaluating the quality of multimedia codecs, this paper instead introduces a framework for granular rating of multiple comparable videos in parallel. This methodology essentially analyses all condition pairs at once. Our contributions are 1) a proposed framework, called HEMVIP, for parallel and granular evaluation of multiple video stimuli and 2) a validation study confirming that results obtained using the tool are in close agreement with results of prior studies using conventional multiple pairwise comparisons.
\end{abstract}



\settopmatter{printfolios=true} 
\maketitle

\section{Introduction}
Owing to the difficulties of objectively quantifying human perception and preference, user studies have become the canonical way to evaluate stimuli in many fields. This includes, for example, aspects of human-computer interaction such as video stimuli of synthetic gesture motion for avatars and social robots \cite{yoon2019robots,alexanderson2020style,jonell2020let,kucherenko2021moving}.
There exist several methods for performing such evaluations, particularly Likert scales \cite{likert1932technique} and pairwise preference tests. These approaches do however not scale well when comparing many different conditions, for example when performing ablation studies or benchmarking a new system against other available approaches.

This paper proposes a novel method for evaluating comparable video stimuli from multiple conditions. Our method is inspired by an evaluation standard called MUSHRA (MUltiple Stimuli with Hidden Reference and Anchor) \cite{itu2015method}, which is widely used for identifying subtle differences in audio quality. We present HEMVIP (Human Evaluation of Multiple Videos in Parallel), which adapts MUSHRA to the evaluation of video stimuli, and validate our proposal by comparing results obtained from HEMVIP against a previous evaluation of videos of generated non-verbal behavior motion for a virtual agent \cite{kucherenko2020gesticulator}.
Our code is available at \href{https://github.com/jonepatr/hemvip/}{https://github.com/jonepatr/hemvip/} and the data and analysis code for the validation study we describe can be found at \href{https://doi.org/10.5281/zenodo.5196581}{https://doi.org/10.5281/zenodo.5196581}.

\section{Related work}
Subjective evaluations of video stimuli are often carried out using Mean Opinion Scores (MOS) \cite{itu1996telephone}, Likert-type scales \cite{likert1932technique,schrum2020four}, or relative preference tests.
In this paper, we use the problem of evaluating nonverbal-behavior-generation systems for embodied conversational agents (ECAs) as a running example, where both MOS (cf.\ \cite{ishii2018generating,yoon2019robots}) and pairwise preference tests (cf.\ \cite{chiu2014gesture,kucherenko2020gesticulator}) are commonplace; see \cite{wolfert2021review} for a comprehensive review.
In MOS tests, participants rate individual stimuli (in our case, videos) on a discrete scale, e.g., 1 through 5.
A proper Likert-scale evaluation requires many such judgments \cite{schrum2020four}.
Since videos are rated in isolation, raters may struggle to notice minor differences between stimuli and to apply a consistent standard to stimuli presented at different points during the test.
Relative preference tests are easier than MOS for participants as selection is easier than scoring \cite{clark2018rate}, and such tests are usually better at identifying subtle differences. However, relative preference tests introduce additional design choices, e.g., how many pairs to be compared and in which combinations. The binary nature of many preference tests means that responses are relatively information-poor and makes it harder to verify that two conditions are statistically different.
Moreover, neither of these two evaluation schemes scales well with the number of systems to be evaluated.

\begin{figure*}[t]
\centering
\includegraphics[width=0.65\textwidth]{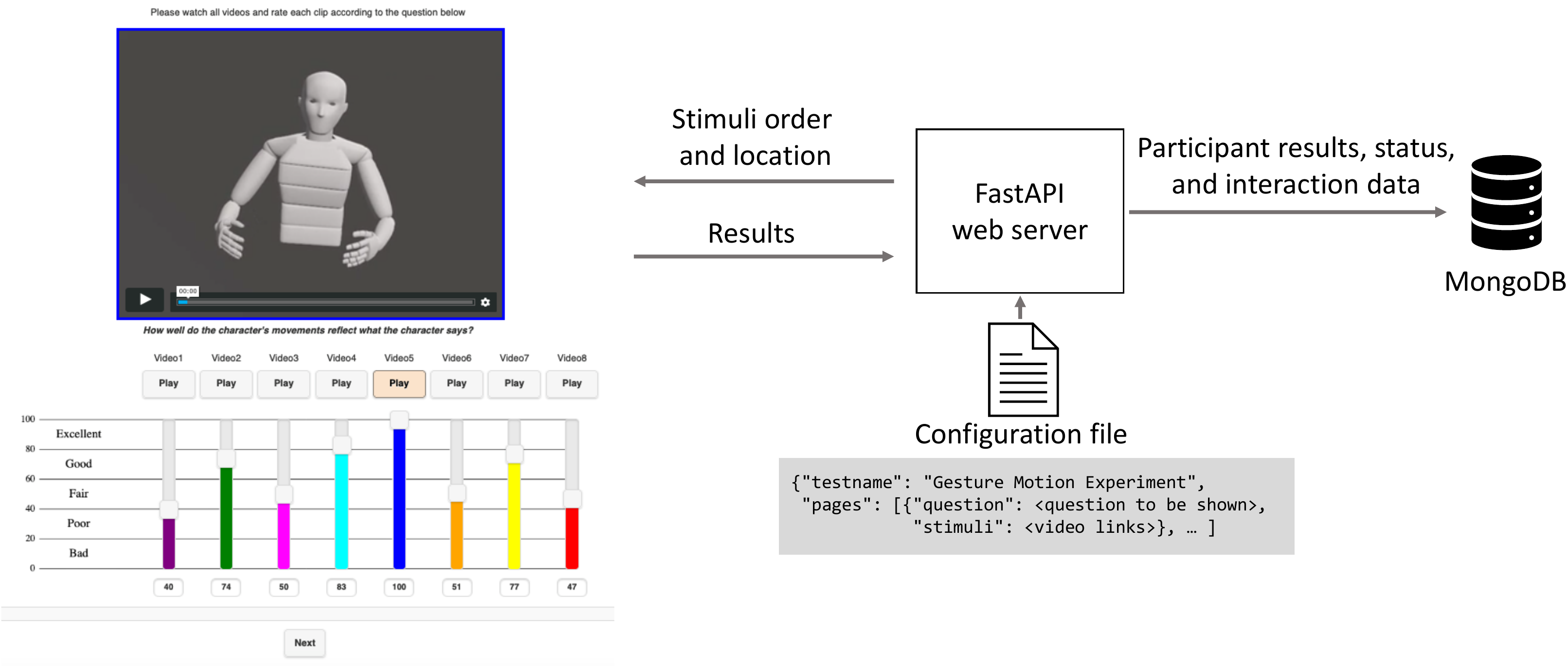}
\vspace{-4mm}
\caption{On the left: A screenshot of a page with stimuli from the HEMVIP evaluation interface. 
On the right: System overview showing how the web server interacts with the evaluation interface.}
\label{fig:evaluation_interace}
\vspace{-3.5mm}
\end{figure*}

The parallel field of audio-quality evaluation has long used both MOS and pairwise preference tests \cite{wester2015using}, but also a more recent standard for comparing multiple audio systems called MUSHRA \cite{itu2015method}.
While originally proposed for comparing audio coding systems, MUSHRA-like tests have also been shown to be more efficient than comparable MOS tests for evaluating speech synthesizers \cite{ribeiro2015perceptual}.
In a MUSHRA test, multiple comparable stimuli (e.g., stimuli generated from the same input text or audio) are presented to the listener on the same page, and listeners rate each stimulus individually in the context of the other stimuli. The main benefit of MUSHRA over the MOS setup is that related stimuli are assessed together, which makes differences easier to spot.
Unlike pairwise tests, many systems can be compared at once.
In contrast to both MOS and preference tests, a high-resolution scale is used, which can resolve smaller differences and use more sensitive statistical analyses.
Recent software for conducting MUSHRA tests online \cite{schoeffler2018webmushra} makes it easy to evaluate synthetic audio using crowdsourcing platforms.

Our proposal also has elements in common with the MUSHRA-derived ITU SAMVIQ (Subjective Assessment Methodology for Video Quality) standard \cite{rec1788bt}, which, however, has been withdrawn in early 2020. 
SAMVIQ was proposed for evaluation of video codec quality and functions similarly to MUSHRA, using hidden references and anchors.
Braude also used a MUSHRA-derived setup for videos, specifically to evaluate head-motion synthesis, however, only one video was shown at a time together with a reference \cite{braude2016head}.
Our proposal instead adapts the MUSHRA approach to the parallel evaluation of video stimuli, to meet the needs of user studies and subjective evaluations in, e.g., human-computer interaction.
\vspace{-0.75mm}

\section{HEMVIP}
\label{sec:HEMVIP}
The HEMVIP framework is an extension of the MUSHRA standard \cite{itu2015method}, but instead of audio recordings we assess multiple comparable videos together using the same setup of parallel rating sliders.
The core aspects of MUSHRA that HEMVIP inherits are 1) joint, parallel scoring of comparable stimuli, e.g., ones intended for the same context or corresponding to the same system input, which makes HEMVIP scale well for multiple comparisons; 2) blind judgments, in that the systems being compared are unlabeled and the order typically is random; and 3) the use of highly granular ratings entered via a slider GUI (0--100 being the default).
There are also some differences from MUSHRA that go beyond the fact that HEMVIP uses video instead of audio:
Unlike audio-codec evaluations, where signal bandwidth can be reduced to degrade quality, there is no canonical way to define objectively poor and degraded stimuli to use as low-end anchor stimuli in, e.g., gesture motion.
HEMVIP thus does not mandate a specific low-end anchor (one can easily be included if it exists), nor does it provide an explicit reference example by default.
Absent this reference, there is no requirement to rate the perceived best stimulus on each page a perfect 100, since there is no assumption that any of the stimuli are perfect.

Fig.~\ref{fig:evaluation_interace} shows an example of the HEMVIP evaluation interface (on the left).
Each ``Play'' button corresponds to a video stimulus, which is played upon clicking the button. Clicking any other play button will immediately start playing its corresponding video instead.
Below the play buttons are sliders for rating the different stimuli in response to the question shown below the video.
In its current incarnation, HEMVIP is best suited to evaluate one question per page.
Text on the left-hand side, here the same labels as used by MOS tests \cite{itu1996telephone}, is used to anchor different intervals on the scale.

We implemented the HEMVIP framework based on webMUSHRA \cite{schoeffler2018webmushra},
but modified to support video material, and including a fully configured web server for the task.
%
%
Configuration files, in JSON, are used to set up an experiment.
These files for instance define the pages (including instructions and a post-test survey) and questions, and which stimuli that are shown.
To allow counterbalancing there is one file per participant, that specifies the order of pages and the stimuli on each page.
To weed out inattentive or non-serious test-takers, we also implemented a mechanism for attention checks where participants were required to input a specific value for a certain stimulus (within a tolerance of plus/minus 3 and accepting acoustically ambiguous numbers such as 13 vs.\ 30). An instruction to set this value was inserted into the attention-check stimulus, either through a text overlay or using a synthesized voice, which otherwise appeared similar to other videos.

During the test, every click in the interface is recorded in an interaction log together with timestamps and what element was clicked. This could potentially be used for interaction analysis or detecting cheating participants. The time a participant spends watching each video is also recorded. To make it easier to remember which slider goes with which video, we apply individual colors to each slider and a colored border around the video, matching the color of the slider of the currently playing stimulus (see Fig.~\ref{fig:evaluation_interace}). These colors were assigned randomly for each page, and test participants were informed that assignments were completely random.

In addition to extending webMUSHRA with additional JavaScript templates and functionality, our HEMVIP implementation incorporates tools to create configuration files and managing participants and a Python-based web server using \href{https://fastapi.tiangolo.com/}{FastAPI} to provide the configuration files to the frontend, block users that fail attention checks, and save result data in an external \href{https://www.mongodb.com/}{MongoDB} database, 
see Fig.~\ref{fig:evaluation_interace}.
\vspace{-4mm}

\section{Validation study}
We validated the results obtained using HEMVIP by reproducing a previously conducted evaluation \cite{kucherenko2020gesticulator}, where different speech gesture generation models were compared. This previous evaluation contained six individual ablation studies and a study comparing the best ablation to the ground truth. Our validation study compared those six ablations (named NoAR, NoPCA, NoFiLM, NoAudio, NoText, and NoVel), the ``Full'' model, and the ground truth side by side (see \href{http://doi.org/10.6084/m9.figshare.13055609}{http://doi.org/10.6084/m9.figshare.13055609} for video examples and \cite{kucherenko2020gesticulator} for the details of the ablated and ``Full'' model).

Participants were asked to rate video stimuli produced by various gesture generation models. The stimuli from all conditions were presented together in parallel, and the participants were asked to rate them individually. The study was balanced such that each stimulus appeared on each page with approximately equal frequency (stimulus order), and each condition was associated with each slider with approximately equal frequency (condition order). For any given participant and study, each page would use different speech segments. Every page would contain the ``Full'' condition, the ground truth condition, and additionally five or six of the ablated conditions, depending on whether an attention check was employed or not. Three attention checks were incorporated into the pages for each study participant as described in Section~\ref{sec:HEMVIP}, using randomly selected numbers between 5 and 95.
Which sliders on which pages that were used for attention checks was uniformly random, except that no page had more than one attention check, and condition ``Full'' and the ground truth never were replaced by attention checks.

Four separate studies were conducted, one for each question. Since the original questions were meant for pairwise comparisons, they were slightly modified to fit the parallel question context, but are still in close agreement with the questions asked in \cite{kucherenko2020gesticulator}:\\
\textbf{Q1} ``In which video are the character's movements most human-like?'' became ``How human-like are the character's movements?'' \\
\textbf{Q2} ``In which video do the character's movements most reflect what the character says?'' became ``How well do the character's movements reflect what the character says?''\\
\textbf{Q3} ``In which video do the character's movements most help to understand what the character says?'' became ``How well do the character's movements help to understand what the character says?''\\
\textbf{Q4} ``In which video are the character's voice and movement more in sync?'' became ``How well are the character's voice and movements in sync?''

After completing the 10 stimuli pages, the participants answered a questionnaire asking demographic questions (age, gender, which continent they had lived on the most, whether English was their native language and how they perceived the task difficulty) together with more qualitative questions not covered in this paper.

The video stimuli were the same as in \cite{kucherenko2020gesticulator}\footnote{Obtained online from \href{http://doi.org/10.6084/m9.figshare.13055609}{http://doi.org/10.6084/m9.figshare.13055609}.}. The clips were 10~s long and comprised generated motion (or ground truth motion capture) together with matching speech from a single actor originally recorded by \cite{ferstl2018investigating}.
There were 50 speech segments in total, each associated with one motion from each of the eight conditions.

We used the crowdsourcing platform \href{https://www.prolific.co}{Prolific} to recruit 46 participants per study (totaling 184 participants). This resulted in approximately as many ratings as in the data from  \cite{kucherenko2020gesticulator}\footnote{Obtained online from \href{https://doi.org/10.6084/m9.figshare.13055585.v6}{https://doi.org/10.6084/m9.figshare.13055585.v6}.}%
, i.e., 4 studies, 46 participants, 10 pages, 8 ratings per page, minus 552 attention checks,\vspace{-0.001ex} for a total of 14,168 ratings, to be compared against 143 participants, 26 pages, 4 ratings per page, minus 858 attention checks, totaling 14,014 ratings. Demographics for each study are shown in Table~\ref{tab:demographics}. 
The participants were paid 6 GBP for completing the study.

\begin{table*}[hbt!]
\centering
\small
\caption{Significant results found by HEMVIP against those found by a re-analysis of the data from the pairwise user study in \cite{kucherenko2020gesticulator}.
``Significant in \cite{kucherenko2020gesticulator}?'' marks which system contrasts that were found to be significantly different by Holm-Bonferroni-corrected Clopper-Pearson (C-P) tests (preference ignoring ties) applied to the data from \cite{kucherenko2020gesticulator}.
Other rows report whether or not these findings of significance agree with the significant differences found by HEVMIP in our user studies, using either C-P tests or pairwise $t$-tests (mean rating), both with Holm-Bonferroni correction.
Empty cells signify agreement (no difference), while $+$ or $-$ mean that HEMVIP found, respectively did not find, the system contrast significant, unlike ``Significant in \cite{kucherenko2020gesticulator}?''.}
\label{tab:overview}
\vspace{-2mm}
\begin{tabular}{@{}l|cccc|cccc|cccc|cccc|cccc|cccc|cccc@{}}
\toprule 
System contrast & 
\multicolumn{4}{c|}{Full${\to}$NoAR} &
\multicolumn{4}{c|}{Full${\to}$NoPCA} &
\multicolumn{4}{c|}{Full${\to}$NoFiLM} & \multicolumn{4}{c|}{Full${\to}$NoAudio} & \multicolumn{4}{c|}{Full${\to}$NoText} &
\multicolumn{4}{c|}{Full${\to}$NoVel} &
\multicolumn{4}{c}{NoPCA${\to}$GT}
\tabularnewline
Study question & 
Q1 & Q2 & Q3 & Q4 & 
Q1 & Q2 & Q3 & Q4 & 
Q1 & Q2 & Q3 & Q4 & 
Q1 & Q2 & Q3 & Q4 & 
Q1 & Q2 & Q3 & Q4 & 
Q1 & Q2 & Q3 & Q4 & 
Q1 & Q2 & Q3 & Q4
\tabularnewline
\midrule 
Significant in \cite{kucherenko2020gesticulator}? & 
\cmark & \xmark & \xmark & \xmark  & 
\cmark & \cmark & \cmark & \cmark  & 
\cmark & \cmark & \cmark & \cmark  & 
\cmark & \cmark & \cmark & \cmark  & 
\cmark & \cmark & \cmark & \cmark  & 
\cmark & \xmark & \cmark & \cmark  & 
\cmark & \cmark & \cmark & \cmark
\tabularnewline

C-P test difference &
&     &     & $\mathbf{+}$ & 
&     &     &              & 
&     &     & $-$          & 
& $-$ &     &              & 
&     &     &              & 
&     & $-$ & $-$          & 
&     &     &
\tabularnewline
$t$-test difference & 
& & & $\mathbf{+}$  & 
& & &               & 
& & &               & 
& & &               & 
& & &               & 
& & &               & 
& & &
\tabularnewline
\bottomrule
\end{tabular}
\vspace{-2mm}
\end{table*}%
Binary preference judgments as in \cite{kucherenko2020gesticulator} can be analyzed using, e.g., Clopper-Pearson (C-P) tests.
The granular nature of HEMVIP responses, however, open up additional analysis options, and
HEMVIP ratings can also be analyzed using pairwise $t$-tests (for differences in true mean rating) and pairwise Wilcoxon signed-rank tests (differences in true median), as used for MUSHRA tests in audio.

Detailed results of Clopper-Pearson tests (HEMVIP and \cite{kucherenko2020gesticulator}) and $t$-tests (HEMVIP only) are reported in Table~\ref{tab:pairs},
with an overview provided in Table~\ref{tab:overview}. Pairwise Wilcoxon signed-rank tests found exactly the same contrasts to be significant as our $t$-tests did, except the Full${\to}$NoVel comparison in Q4.
\vspace{-4mm}

\section{Discussion}
If we ignore the magnitude of pairwise differences in HEVMIP responses for the same page, and consider only their sign, we can use C-P tests to analyze preference ignoring ties seen in the data from \cite{kucherenko2020gesticulator} and in our ratings.
A comparison shows a high, but not perfect, correspondence between the two evaluations (23 out of 28 contrasts agree on whether or not a difference is significant). The 5 differing contrasts do however show the same direction of preference.
Among analysis methods that also leverage the granular nature of HEMVIP ratings (and which cannot be applied to preference tests like \cite{kucherenko2020gesticulator}), both pairwise $t$-tests and Wilcoxon signed-rank tests produce results that match the original study for all contrasts, except one for the $t$-tests and two for the Wilcoxon signed-rank tests (see Table~\ref{tab:overview} for an overview or
Table~\ref{tab:pairs} for detailed results, confidence intervals, and $p$-values). This validates that the new methodology delivers conclusions highly similar to the conventional pairwise evaluation.

The fact that C-P tests identified fewer significant differences than $t$-tests and Wilcoxon signed-rank tests did on the HEMVIP responses is likely because C-P tests only consider the sign of the difference between paired system ratings.
This uses less of the information available in the ratings, thus requiring more samples to identify a statistically significant difference.
We therefore recommend the two more granular tests for HEMVIP (especially the Wilcoxon, which does not assume Gaussianity) as they leverage more of the information in the responses and are better at telling conditions apart.
HEMVIP additionally allows for comparing all systems against one another (28 pairs), and not only the contrasts in Table~\ref{tab:overview}, at no additional cost. Analyzing our data in this way using the Wilcoxon test finds (after Holm-Bonferroni correction) 26, 23, 24, and 24 of the 28 pairs to be statistically significantly different from one another at $\alpha=0.05$, for questions Q1, Q2, Q3, and Q4.

One of the key benefits of HEMVIP is its efficiency, i.e., the time it takes to provide a certain amount of ratings for a single question. Compared to pairwise comparisons like in, e.g., \cite{kucherenko2020gesticulator,jonell2020iva_crowd}, HEMVIP is more efficient as it evaluates multiple stimuli in parallel. 
To quantify this efficiency, it is perhaps more meaningful to compare against the partial replication study in \cite{jonell2020iva_crowd}. That study only considered the contrast ``NoPCA'' vs.\ ``NoText'' but otherwise used the same evaluation methodology as \cite{kucherenko2020gesticulator} and reached the same conclusions.
Although \cite{kucherenko2020gesticulator} evaluated a greater set of conditions than \cite{jonell2020iva_crowd} (which made it more relevant for our validation study, by providing more data on how HEMVIP compares to a pairwise approach) it is not as suitable for comparing test-taking efficiency to HEMVIP.
This is because \cite{kucherenko2020gesticulator} asked four questions simultaneously, whereas HEMVIP and \cite{jonell2020iva_crowd} present one question at a time, and \cite{kucherenko2020gesticulator} was conducted on \href{https://www.mturk.com}{Amazon Mechanical Turk}, while \cite{jonell2020iva_crowd} also considered Prolific, like the validation study here. (Also, the attention checks in \cite{kucherenko2020gesticulator} were based on detecting heavily degraded videos and were more subtle than the attention checks employed here, but this is less relevant to timing information.)
Prolific participants in \cite{jonell2020iva_crowd} took approximately 32 s to complete one page in a pairwise comparison, while the average time our 
study participants spent per page was 176 s. This means that comparing one system to the other seven in a pairwise fashion takes 224 s (7 $\times$ 32 s), as opposed to 176 s using HEMVIP. Furthermore, if one wants to compare all systems with each other, HEMVIP still only needs 176 s (one page), as opposed to 896 s (8 choose 2 = 28 pages) using a pairwise method, where each system would be compared against a single other system on each page.
HEMVIP thus scales much better than pairwise evaluations.



A limitation of HEMVIP is that it mainly focuses on one question at a time. A way of scoring multiple questions at once (e.g., for efficiently conducting proper Likert-scale evaluations; cf.\ \cite{schrum2020four}) could probably be added by, for example, adding a new set of sliders after one question has been rated, or putting in multiple sliders per stimulus. This -- and related issues such as how many stimuli that can be evaluated in parallel without exhausting the participant -- 
should be addressed in a future usability study, as it is currently unknown if adding an extra question or increasing the amount of stimuli would be too cognitively\vspace{-0.001ex} demanding for the users. The MUSHRA standard for audio \cite{itu2015method} recommends no more than 12 stimuli 
per page. 
A majority of participants indicated repetitiveness from rating 8 videos with the same sound per page, which could indicate that a lower number than 8 could be preferable. Another limitation of this work is that we only evaluated video stimuli from one particular domain and task. While we do evaluate it using material that has been evaluated twice before, and found consistent results, future work should investigate different kinds of video stimuli to further study how well the method generalizes to different situations.
\vspace{-0.5mm}

\section{Conclusions and future work}
In this paper we proposed a framework for evaluating multiple comparable video stimuli in parallel, called HEMVIP. A validation experiment compared results obtained using the proposed method to results obtained using pairwise binary preference tests in earlier work in the domain of nonverbal behavior, finding high correspondence between the two experiments, but with greater efficiency and vastly better scaling properties for HEMVIP. It has for example already been used for a large-scale evaluation in the GENEA Challenge 2020 \cite{kucherenko2021large}.
We believe HEMVIP can be of great benefit to researchers performing thorough evaluations across multiple video stimuli -- not only as an alternative to pairwise video presentations but also as a replacement for many MOS tests \cite{itu1996telephone}. 
Furthermore, the video stimuli need not show or compare gestures or feature ECAs: the key point is that stimuli from all conditions can be compared side by side, e.g., different voiceover audio for the same video clip, or comparisons of different signal processing methods. With some code modifications, stimuli may also be displayed on a separate device, e.g., for VR/AR evaluations.

Future work includes validating
on stimuli from other domains,
along with usability testing of various aspects of the interface, such as 
how many videos users are able to rate effectively on a single page without incurring excessive cognitive load.

\vspace{-0.5mm}
\begin{acks}
The authors wish to thank Jonas Beskow, Dmytro Kalpakchi, and Bram Willemsen for their feedback on the paper.

This research was partially supported by Swedish Foundation for Strategic Research contract no.\ RIT15-0107 (EACare), by IITP grant no.\ 2017-0-00162 (Development of Human-care Robot Technology for Aging Society) funded by the Korean government (MSIT), by the Flemish Research Foundation grant no.\ 1S95020N, and by the Wallenberg AI, Autonomous Systems and Software Program (WASP) funded by the Knut and Alice Wallenberg Foundation.
\end{acks}
\balance
\bibliographystyle{ACM-Reference-Format}
\bibliography{ref}


\begin{thebibliography}{21}


\ifx \showCODEN    \undefined \def \showCODEN     #1{\unskip}     \fi
\ifx \showDOI      \undefined \def \showDOI       #1{#1}\fi
\ifx \showISBNx    \undefined \def \showISBNx     #1{\unskip}     \fi
\ifx \showISBNxiii \undefined \def \showISBNxiii  #1{\unskip}     \fi
\ifx \showISSN     \undefined \def \showISSN      #1{\unskip}     \fi
\ifx \showLCCN     \undefined \def \showLCCN      #1{\unskip}     \fi
\ifx \shownote     \undefined \def \shownote      #1{#1}          \fi
\ifx \showarticletitle \undefined \def \showarticletitle #1{#1}   \fi
\ifx \showURL      \undefined \def \showURL       {\relax}        \fi
\providecommand\bibfield[2]{#2}
\providecommand\bibinfo[2]{#2}
\providecommand\natexlab[1]{#1}
\providecommand\showeprint[2][]{arXiv:#2}

\bibitem[\protect\citeauthoryear{Alexanderson, Henter, Kucherenko, and
  Beskow}{Alexanderson et~al\mbox{.}}{2020}]%
        {alexanderson2020style}
\bibfield{author}{\bibinfo{person}{Simon Alexanderson},
  \bibinfo{person}{Gustav~Eje Henter}, \bibinfo{person}{Taras Kucherenko},
  {and} \bibinfo{person}{Jonas Beskow}.} \bibinfo{year}{2020}\natexlab{}.
\newblock \showarticletitle{Style-controllable speech-driven gesture synthesis
  using normalising flows}.
\newblock \bibinfo{journal}{\emph{Computer Graphics Forum}}
  \bibinfo{volume}{39}, \bibinfo{number}{2} (\bibinfo{year}{2020}),
  \bibinfo{pages}{487--496}.
\newblock
\urldef\tempurl%
\url{https://doi.org/10.1111/cgf.13946}
\showDOI{\tempurl}


\bibitem[\protect\citeauthoryear{Braude}{Braude}{2016}]%
        {braude2016head}
\bibfield{author}{\bibinfo{person}{David~A. Braude}.}
  \bibinfo{year}{2016}\natexlab{}.
\newblock \emph{\bibinfo{title}{Head motion synthesis: Evaluation and a
  template motion approach}}.
\newblock \bibinfo{thesistype}{Ph.D. Dissertation}. \bibinfo{school}{The
  University of Edinburgh}, \bibinfo{address}{UK}.
\newblock
\urldef\tempurl%
\url{http://hdl.handle.net/1842/20418}
\showURL{%
\tempurl}


\bibitem[\protect\citeauthoryear{Chiu and Marsella}{Chiu and Marsella}{2014}]%
        {chiu2014gesture}
\bibfield{author}{\bibinfo{person}{Chung-Cheng Chiu} {and}
  \bibinfo{person}{Stacy Marsella}.} \bibinfo{year}{2014}\natexlab{}.
\newblock \showarticletitle{Gesture generation with low-dimensional
  embeddings}. In \bibinfo{booktitle}{\emph{Proceedings of the International
  Conference on Autonomous Agents and Multi-agent Systems}}.
  \bibinfo{publisher}{ACM}, \bibinfo{pages}{781--788}.
\newblock
\urldef\tempurl%
\url{https://doi.org/10.5555/2615731.2615857}
\showDOI{\tempurl}


\bibitem[\protect\citeauthoryear{Clark, Howard, Woods, Penton-Voak, and
  Neumann}{Clark et~al\mbox{.}}{2018}]%
        {clark2018rate}
\bibfield{author}{\bibinfo{person}{Andrew~P. Clark}, \bibinfo{person}{Kate~L.
  Howard}, \bibinfo{person}{Andy~T. Woods}, \bibinfo{person}{Ian~S.
  Penton-Voak}, {and} \bibinfo{person}{Christof Neumann}.}
  \bibinfo{year}{2018}\natexlab{}.
\newblock \showarticletitle{Why rate when you could compare? Using the
  ``EloChoice'' package to assess pairwise comparisons of perceived physical
  strength}.
\newblock \bibinfo{journal}{\emph{PLOS One}} \bibinfo{volume}{13},
  \bibinfo{number}{1} (\bibinfo{year}{2018}), \bibinfo{pages}{1--16}.
\newblock
\urldef\tempurl%
\url{https://doi.org/10.1371/journal.pone.0190393}
\showDOI{\tempurl}


\bibitem[\protect\citeauthoryear{Ferstl and McDonnell}{Ferstl and
  McDonnell}{2018}]%
        {ferstl2018investigating}
\bibfield{author}{\bibinfo{person}{Ylva Ferstl} {and} \bibinfo{person}{Rachel
  McDonnell}.} \bibinfo{year}{2018}\natexlab{}.
\newblock \showarticletitle{Investigating the use of recurrent motion modelling
  for speech gesture generation}. In \bibinfo{booktitle}{\emph{Proceedings of
  the ACM International Conference on Intelligent Virtual Agents}}.
  \bibinfo{publisher}{ACM}, \bibinfo{pages}{93--98}.
\newblock
\urldef\tempurl%
\url{https://doi.org/10.1145/3267851.3267898}
\showDOI{\tempurl}


\bibitem[\protect\citeauthoryear{Ishii, Katayama, Higashinaka, and
  Tomita}{Ishii et~al\mbox{.}}{2018}]%
        {ishii2018generating}
\bibfield{author}{\bibinfo{person}{Ryo Ishii}, \bibinfo{person}{Taichi
  Katayama}, \bibinfo{person}{Ryuichiro Higashinaka}, {and}
  \bibinfo{person}{Junji Tomita}.} \bibinfo{year}{2018}\natexlab{}.
\newblock \showarticletitle{Generating body motions using spoken language in
  dialogue}. In \bibinfo{booktitle}{\emph{Proceedings of the ACM International
  Conference on Intelligent Virtual Agents}}. \bibinfo{publisher}{ACM},
  \bibinfo{pages}{87--92}.
\newblock
\urldef\tempurl%
\url{https://doi.org/10.1145/3267851.3267866}
\showDOI{\tempurl}


\bibitem[\protect\citeauthoryear{{ITU-R BS.1534-3}}{{ITU-R BS.1534-3}}{2015}]%
        {itu2015method}
\bibfield{author}{\bibinfo{person}{{ITU-R BS.1534-3}}.}
  \bibinfo{year}{2015}\natexlab{}.
\newblock \bibinfo{booktitle}{\emph{Method for the Subjective Assessment of
  Intermediate Quality Level of Audio Systems}}.
\newblock \bibinfo{type}{Standard}. \bibinfo{institution}{ITU}.
\newblock
\urldef\tempurl%
\url{https://www.itu.int/rec/R-REC-BS.1534-3-201510-I}
\showURL{%
\tempurl}


\bibitem[\protect\citeauthoryear{{ITU-R BT.1788}}{{ITU-R BT.1788}}{2007}]%
        {rec1788bt}
\bibfield{author}{\bibinfo{person}{{ITU-R BT.1788}}.}
  \bibinfo{year}{2007}\natexlab{}.
\newblock \bibinfo{booktitle}{\emph{Methodology for the Subjective Assessment
  of Video Quality in Multimedia Applications}}.
\newblock \bibinfo{type}{Standard}. \bibinfo{institution}{ITU}.
\newblock
\urldef\tempurl%
\url{https://www.itu.int/rec/R-REC-BT.1788}
\showURL{%
\tempurl}


\bibitem[\protect\citeauthoryear{{ITU-T P.800}}{{ITU-T P.800}}{1996}]%
        {itu1996telephone}
\bibfield{author}{\bibinfo{person}{{ITU-T P.800}}.}
  \bibinfo{year}{1996}\natexlab{}.
\newblock \bibinfo{booktitle}{\emph{Methods for Subjective Determination of
  Transmission Quality}}.
\newblock \bibinfo{type}{Standard}. \bibinfo{institution}{ITU}.
\newblock
\urldef\tempurl%
\url{https://www.itu.int/rec/T-REC-P.800-199608-I}
\showURL{%
\tempurl}


\bibitem[\protect\citeauthoryear{Jonell, Kucherenko, Henter, and Beskow}{Jonell
  et~al\mbox{.}}{2020a}]%
        {jonell2020let}
\bibfield{author}{\bibinfo{person}{Patrik Jonell}, \bibinfo{person}{Taras
  Kucherenko}, \bibinfo{person}{Gustav~Eje Henter}, {and}
  \bibinfo{person}{Jonas Beskow}.} \bibinfo{year}{2020}\natexlab{a}.
\newblock \showarticletitle{Let's face it: Probabilistic multi-modal
  interlocutor-aware generation of facial gestures in dyadic settings}. In
  \bibinfo{booktitle}{\emph{Proceedings of the ACM International Conference on
  Intelligent Virtual Agents}}. \bibinfo{publisher}{ACM}, Article
  \bibinfo{articleno}{31}, \bibinfo{numpages}{8}~pages.
\newblock
\urldef\tempurl%
\url{https://doi.org/10.1145/3383652.3423911}
\showDOI{\tempurl}


\bibitem[\protect\citeauthoryear{Jonell, Kucherenko, Torre, and Beskow}{Jonell
  et~al\mbox{.}}{2020b}]%
        {jonell2020iva_crowd}
\bibfield{author}{\bibinfo{person}{Patrik Jonell}, \bibinfo{person}{Taras
  Kucherenko}, \bibinfo{person}{Ilaria Torre}, {and} \bibinfo{person}{Jonas
  Beskow}.} \bibinfo{year}{2020}\natexlab{b}.
\newblock \showarticletitle{Can we trust online crowdworkers? Comparing online
  and offline participants in a preference test of virtual agents.}. In
  \bibinfo{booktitle}{\emph{Proceedings of the ACM International Conference on
  Intelligent Virtual Agents}}. \bibinfo{publisher}{ACM}, Article
  \bibinfo{articleno}{31}, \bibinfo{numpages}{8}~pages.
\newblock
\urldef\tempurl%
\url{https://doi.org/10.1145/3383652.3423860}
\showDOI{\tempurl}


\bibitem[\protect\citeauthoryear{Kucherenko, Hasegawa, Kaneko, Henter, and
  Kjellstr{\"o}m}{Kucherenko et~al\mbox{.}}{2021a}]%
        {kucherenko2021moving}
\bibfield{author}{\bibinfo{person}{Taras Kucherenko}, \bibinfo{person}{Dai
  Hasegawa}, \bibinfo{person}{Naoshi Kaneko}, \bibinfo{person}{Gustav~Eje
  Henter}, {and} \bibinfo{person}{Hedvig Kjellstr{\"o}m}.}
  \bibinfo{year}{2021}\natexlab{a}.
\newblock \showarticletitle{Moving fast and slow: Analysis of representations
  and post-processing in speech-driven automatic gesture generation}.
\newblock \bibinfo{journal}{\emph{International Journal of Human–Computer
  Interaction}} \bibinfo{volume}{37}, \bibinfo{number}{14}
  (\bibinfo{year}{2021}), \bibinfo{pages}{1300--1316}.
\newblock
\urldef\tempurl%
\url{https://doi.org/10.1080/10447318.2021.1883883}
\showDOI{\tempurl}


\bibitem[\protect\citeauthoryear{Kucherenko, Jonell, van Waveren, Henter,
  Alexanderson, Leite, and Kjellstr{\"o}m}{Kucherenko et~al\mbox{.}}{2020}]%
        {kucherenko2020gesticulator}
\bibfield{author}{\bibinfo{person}{Taras Kucherenko}, \bibinfo{person}{Patrik
  Jonell}, \bibinfo{person}{Sanne van Waveren}, \bibinfo{person}{Gustav~Eje
  Henter}, \bibinfo{person}{Simon Alexanderson}, \bibinfo{person}{Iolanda
  Leite}, {and} \bibinfo{person}{Hedvig Kjellstr{\"o}m}.}
  \bibinfo{year}{2020}\natexlab{}.
\newblock \showarticletitle{Gesticulator: A framework for semantically-aware
  speech-driven gesture generation}. In \bibinfo{booktitle}{\emph{Proceedings
  of the ACM International Conference on Multimodal Interaction}}.
  \bibinfo{publisher}{ACM}, \bibinfo{pages}{242--250}.
\newblock
\urldef\tempurl%
\url{https://doi.org/10.1145/3382507.3418815}
\showDOI{\tempurl}


\bibitem[\protect\citeauthoryear{Kucherenko, Jonell, Yoon, Wolfert, and
  Henter}{Kucherenko et~al\mbox{.}}{2021b}]%
        {kucherenko2021large}
\bibfield{author}{\bibinfo{person}{Taras Kucherenko}, \bibinfo{person}{Patrik
  Jonell}, \bibinfo{person}{Youngwoo Yoon}, \bibinfo{person}{Pieter Wolfert},
  {and} \bibinfo{person}{Gustav~Eje Henter}.} \bibinfo{year}{2021}\natexlab{b}.
\newblock \showarticletitle{A large, crowdsourced evaluation of gesture
  generation systems on common data: The {GENEA} {C}hallenge 2020}. In
  \bibinfo{booktitle}{\emph{Proceedings of the International Conference on
  Intelligent User Interfaces}}. \bibinfo{publisher}{ACM},
  \bibinfo{pages}{11--21}.
\newblock
\urldef\tempurl%
\url{https://doi.org/10.1145/3397481.3450692}
\showDOI{\tempurl}


\bibitem[\protect\citeauthoryear{Likert}{Likert}{1932}]%
        {likert1932technique}
\bibfield{author}{\bibinfo{person}{Rensis Likert}.}
  \bibinfo{year}{1932}\natexlab{}.
\newblock \showarticletitle{A technique for the measurement of attitudes}.
\newblock \bibinfo{journal}{\emph{Archives of Psychology}}
  \bibinfo{volume}{140} (\bibinfo{year}{1932}), \bibinfo{pages}{1--55}.
\newblock


\bibitem[\protect\citeauthoryear{Ribeiro, Yamagishi, and Clark}{Ribeiro
  et~al\mbox{.}}{2015}]%
        {ribeiro2015perceptual}
\bibfield{author}{\bibinfo{person}{Manuel~Sam Ribeiro},
  \bibinfo{person}{Junichi Yamagishi}, {and} \bibinfo{person}{Robert A.~J.
  Clark}.} \bibinfo{year}{2015}\natexlab{}.
\newblock \showarticletitle{A perceptual investigation of wavelet-based
  decomposition of $f$0 for text-to-speech synthesis}. In
  \bibinfo{booktitle}{\emph{Proceedings of the Annual Conference of the
  International Speech Communication Association}}. \bibinfo{publisher}{ISCA},
  \bibinfo{pages}{1586--1590}.
\newblock
\urldef\tempurl%
\url{https://doi.org/10.21437/Interspeech.2015-368}
\showDOI{\tempurl}


\bibitem[\protect\citeauthoryear{Schoeffler, Bartoschek, St{\"o}ter, Roess,
  Westphal, Edler, and Herre}{Schoeffler et~al\mbox{.}}{2018}]%
        {schoeffler2018webmushra}
\bibfield{author}{\bibinfo{person}{Michael Schoeffler}, \bibinfo{person}{Sarah
  Bartoschek}, \bibinfo{person}{Fabian-Robert St{\"o}ter},
  \bibinfo{person}{Marlene Roess}, \bibinfo{person}{Susanne Westphal},
  \bibinfo{person}{Bernd Edler}, {and} \bibinfo{person}{J{\"u}rgen Herre}.}
  \bibinfo{year}{2018}\natexlab{}.
\newblock \showarticletitle{webMUSHRA --- A comprehensive framework for
  web-based listening tests}.
\newblock \bibinfo{journal}{\emph{Journal of Open Research Software}}
  \bibinfo{volume}{6}, \bibinfo{number}{1} (\bibinfo{year}{2018}),
  \bibinfo{pages}{8:1--8:8}.
\newblock
\urldef\tempurl%
\url{https://doi.org/10.5334/jors.187}
\showDOI{\tempurl}


\bibitem[\protect\citeauthoryear{Schrum, Johnson, Ghuy, and Gombolay}{Schrum
  et~al\mbox{.}}{2020}]%
        {schrum2020four}
\bibfield{author}{\bibinfo{person}{Mariah~L. Schrum}, \bibinfo{person}{Michael
  Johnson}, \bibinfo{person}{Muyleng Ghuy}, {and} \bibinfo{person}{Matthew~C.
  Gombolay}.} \bibinfo{year}{2020}\natexlab{}.
\newblock \showarticletitle{Four years in review: Statistical practices of
  Likert scales in human-robot interaction studies}. In
  \bibinfo{booktitle}{\emph{Companion of the 2020 ACM/IEEE International
  Conference on Human-Robot Interaction}}. \bibinfo{publisher}{ACM},
  \bibinfo{pages}{43--52}.
\newblock
\urldef\tempurl%
\url{https://doi.org/10.1145/3371382.3380739}
\showDOI{\tempurl}


\bibitem[\protect\citeauthoryear{Wester, Valentini-Botinhao, and Henter}{Wester
  et~al\mbox{.}}{2015}]%
        {wester2015using}
\bibfield{author}{\bibinfo{person}{Mirjam Wester}, \bibinfo{person}{Cassia
  Valentini-Botinhao}, {and} \bibinfo{person}{Gustav~Eje Henter}.}
  \bibinfo{year}{2015}\natexlab{}.
\newblock \showarticletitle{Are we using enough listeners? {N}o! {A}n
  empirically-supported critique of {I}nterspeech 2014 {TTS} evaluations}. In
  \bibinfo{booktitle}{\emph{Proceedings of the Annual Conference of the
  International Speech Communication Association}}, Vol.~\bibinfo{volume}{16}.
  \bibinfo{publisher}{ISCA}, \bibinfo{pages}{3476--3480}.
\newblock
\urldef\tempurl%
\url{https://doi.org/10.21437/Interspeech.2015-689}
\showDOI{\tempurl}


\bibitem[\protect\citeauthoryear{Wolfert, Robinson, and Belpaeme}{Wolfert
  et~al\mbox{.}}{2021}]%
        {wolfert2021review}
\bibfield{author}{\bibinfo{person}{Pieter Wolfert}, \bibinfo{person}{Nicole
  Robinson}, {and} \bibinfo{person}{Tony Belpaeme}.}
  \bibinfo{year}{2021}\natexlab{}.
\newblock \bibinfo{title}{A review of evaluation practices of gesture
  generation in embodied conversational agents}.
\newblock
\newblock
\showeprint[arxiv]{2101.03769}


\bibitem[\protect\citeauthoryear{Yoon, Ko, Jang, Lee, Kim, and Lee}{Yoon
  et~al\mbox{.}}{2019}]%
        {yoon2019robots}
\bibfield{author}{\bibinfo{person}{Youngwoo Yoon}, \bibinfo{person}{Woo-Ri Ko},
  \bibinfo{person}{Minsu Jang}, \bibinfo{person}{Jaeyeon Lee},
  \bibinfo{person}{Jaehong Kim}, {and} \bibinfo{person}{Geehyuk Lee}.}
  \bibinfo{year}{2019}\natexlab{}.
\newblock \showarticletitle{Robots learn social skills: End-to-end learning of
  co-speech gesture generation for humanoid robots}. In
  \bibinfo{booktitle}{\emph{Proceedings of the IEEE International Conference on
  Robotics and Automation}}. \bibinfo{publisher}{IEEE},
  \bibinfo{pages}{4303--4309}.
\newblock
\urldef\tempurl%
\url{https://doi.org/10.1109/ICRA.2019.8793720}
\showDOI{\tempurl}


\end{thebibliography}


\begin{table*}
\vspace{5mm}
\centering
\caption{Demographics for the four studies.}
\label{tab:demographics}
\begin{tabular}{@{}lcccc@{}}
\toprule 
\multicolumn{1}{l}{Demographic} & Q1 & Q2 & Q3 & Q4 \\ 
 \midrule
\multicolumn{1}{l}{Age} & \multicolumn{1}{c}{$34\pm12$} & \multicolumn{1}{c}{$28\pm9$} & \multicolumn{1}{c}{$31\pm11$} & \multicolumn{1}{c}{$29\pm10$} \tabularnewline

\multicolumn{1}{l}{Gender (F/M)} & \multicolumn{1}{c}{17/29} & \multicolumn{1}{c}{34/12} & \multicolumn{1}{c}{24/22} & \multicolumn{1}{c}{25/21} \tabularnewline

\multicolumn{1}{l}{Native English speakers} & \multicolumn{1}{c}{89\%} & \multicolumn{1}{c}{83\%} & \multicolumn{1}{c}{79\%} & \multicolumn{1}{c}{87\%} \tabularnewline

\multicolumn{1}{l}{Task difficulty\textsuperscript{1}} & \multicolumn{1}{c}{$3.4\pm1.2$} & \multicolumn{1}{c}{$3.6\pm1.2$} & \multicolumn{1}{c}{$3.8\pm1$} & \multicolumn{1}{c}{$3.7\pm1$} \tabularnewline

\multicolumn{1}{l}{Primary residency\textsuperscript{2}} & \multicolumn{1}{c}{1/1/39/5/0/0} & \multicolumn{1}{c}{1/2/35/5/2/1} & \multicolumn{1}{c}{2/2/34/7/1/0} & \multicolumn{1}{c}{2/1/28/14/1/0} 
\tabularnewline
\bottomrule
\multicolumn{5}{l}{\textsuperscript{1}Question: ``This task was really easy'' (1 being disagree and 5 agree)}
\tabularnewline
\multicolumn{5}{l}{\textsuperscript{2}Africa/Asia/Europe/North America/Oceania/South America}
\end{tabular}
\vspace{20mm}
\end{table*}
\begin{table*}
\centering
\caption{System contrasts from user studies, with 95\% confidence intervals and Holm-Bonferroni corrected $p$-values from Clopper-Pearson (C-P) tests (preference ignoring ties) and pairwise $t$-tests (mean rating). $R$ stands for ``rating''. Daggers (\textdagger) mark non-significant differences. The ``Vs.\ \cite{kucherenko2020gesticulator}'' columns report whether or not the findings of significance agree with the C-P tests on data from \cite{kucherenko2020gesticulator} reported in the final two columns. Empty cells signify agreement, while $+$ (or $-$) mean that HEMVIP found (or did not find) the difference significant, in contrast to \cite{kucherenko2020gesticulator}.}
\label{tab:pairs}
\begin{tabular}{@{}ll|ccc|ccc|lc@{}}
\toprule
& $A{\to}B$ & $\mathbb{E}[R_B{-}R_A]$ & $p$-val. & Vs.\ \cite{kucherenko2020gesticulator} &  $\mathbb{P}(R_B{>}R_A)$ & $p$-val. & Vs.\ \cite{kucherenko2020gesticulator} &  $ \mathbb{P}(R_B{>}R_A)$ from \cite{kucherenko2020gesticulator} & $p$-val.   \\
\midrule
 & Full${\to}$NoAR  & ${-}$12.1$\pm$2.4  & <10\textsuperscript{${-}$20}\hphantom{\textsuperscript{\textdagger}}  &  & 27\%{}$\in$(22,32)  & <10\textsuperscript{${-}$20}\hphantom{\textsuperscript{\textdagger}}  &  & 23\%{}$\in$(19,28)  & <10\textsuperscript{${-}$25}\hphantom{\textsuperscript{\textdagger}} \tabularnewline
 & Full${\to}$NoPCA  & \hphantom{${-}$}10.7$\pm$1.9  & <10\textsuperscript{${-}$24}\hphantom{\textsuperscript{\textdagger}}  &  & 76\%{}$\in$(71,80)  & <10\textsuperscript{${-}$25}\hphantom{\textsuperscript{\textdagger}}  &  & 56\%{}$\in$(49,63)  & 0.050\hphantom{\textsuperscript{\textdagger}} \tabularnewline
 & Full${\to}$NoFiLM  & \hphantom{${-}$0}3.5$\pm$1.7  & <10\textsuperscript{${-}$3\hphantom{0}}\hphantom{\textsuperscript{\textdagger}}  &  & 58\%{}$\in$(52,63)  & 0.003\hphantom{\textsuperscript{\textdagger}}  &  & 57\%{}$\in$(50,64)  & 0.049\hphantom{\textsuperscript{\textdagger}} \tabularnewline
Q1 & Full${\to}$NoAudio  & \hphantom{0}${-}$2.9$\pm$1.6  & <10\textsuperscript{${-}$3\hphantom{0}}\hphantom{\textsuperscript{\textdagger}}  &  & 43\%{}$\in$(37,48)  & 0.003\hphantom{\textsuperscript{\textdagger}}  &  & 34\%{}$\in$(28,42)  & <10\textsuperscript{${-}$4\hphantom{0}}\hphantom{\textsuperscript{\textdagger}} \tabularnewline
 & Full${\to}$NoText  & ${-}$11.4$\pm$1.6  & <10\textsuperscript{${-}$36}\hphantom{\textsuperscript{\textdagger}}  &  & 19\%{}$\in$(14,23)  & <10\textsuperscript{${-}$38}\hphantom{\textsuperscript{\textdagger}}  &  & 17\%{}$\in$(12,23)  & <10\textsuperscript{${-}$26}\hphantom{\textsuperscript{\textdagger}} \tabularnewline
 & Full${\to}$NoVel  & \hphantom{${-}$0}3.0$\pm$1.6  & <10\textsuperscript{${-}$3\hphantom{0}}\hphantom{\textsuperscript{\textdagger}}  &  & 60\%{}$\in$(54,65)  & <10\textsuperscript{${-}$3\hphantom{0}}\hphantom{\textsuperscript{\textdagger}}  &  & 59\%{}$\in$(52,66)  & 0.024\hphantom{\textsuperscript{\textdagger}} \tabularnewline
 & NoPCA${\to}$GT  & \hphantom{${-}$}40.5$\pm$1.9  & <10\textsuperscript{${-}$99}\hphantom{\textsuperscript{\textdagger}}  &  & 99\%{}$\in$(97,100)  & <10\textsuperscript{${-}$99}\hphantom{\textsuperscript{\textdagger}}  &  & 97\%{}$\in$(94,99)  & <10\textsuperscript{${-}$85}\hphantom{\textsuperscript{\textdagger}} \tabularnewline
\midrule 
 & Full${\to}$NoAR  & \hphantom{0}${-}$1.1$\pm$2.1  & 0.516\textsuperscript{\textdagger}  &  & 50\%{}$\in$(45,56)  & 0.846\textsuperscript{\textdagger}  &  & 49\%{}$\in$(43,55)  & 0.747\textsuperscript{\textdagger} \tabularnewline
 & Full${\to}$NoPCA  & \hphantom{${-}$}11.5$\pm$1.9  & <10\textsuperscript{${-}$27}\hphantom{\textsuperscript{\textdagger}}  &  & 75\%{}$\in$(70,79)  & <10\textsuperscript{${-}$23}\hphantom{\textsuperscript{\textdagger}}  &  & 69\%{}$\in$(63,75)  & <10\textsuperscript{${-}$10}\hphantom{\textsuperscript{\textdagger}} \tabularnewline
 & Full${\to}$NoFiLM  & \hphantom{${-}$0}2.6$\pm$1.7  & 0.006\hphantom{\textsuperscript{\textdagger}}  &  & 56\%{}$\in$(51,62)  & 0.042\hphantom{\textsuperscript{\textdagger}}  &  & 57\%{}$\in$(51,63)  & 0.039\hphantom{\textsuperscript{\textdagger}} \tabularnewline
Q2  & Full${\to}$NoAudio  & \hphantom{0}${-}$2.9$\pm$1.7  & 0.002\hphantom{\textsuperscript{\textdagger}}  &  & 45\%{}$\in$(40,50)  & 0.119\textsuperscript{\textdagger}  & $-$ & 33\%{}$\in$(27,40)  & <10\textsuperscript{${-}$6\hphantom{0}}\hphantom{\textsuperscript{\textdagger}} \tabularnewline
 & Full${\to}$NoText  & ${-}$12.0$\pm$1.7  & <10\textsuperscript{${-}$37}\hphantom{\textsuperscript{\textdagger}}  &  & 23\%{}$\in$(19,28)  & <10\textsuperscript{${-}$27}\hphantom{\textsuperscript{\textdagger}}  &  & 10\%{}$\in$(\hphantom{0}7,15)  & <10\textsuperscript{${-}$49}\hphantom{\textsuperscript{\textdagger}} \tabularnewline
 & Full${\to}$NoVel  & \hphantom{${-}$0}1.0$\pm$1.7  & 0.516\textsuperscript{\textdagger}  &  & 52\%{}$\in$(47,58)  & 0.761\textsuperscript{\textdagger}  &  & 57\%{}$\in$(50,63)  & 0.067\textsuperscript{\textdagger} \tabularnewline
 & NoPCA${\to}$GT  & \hphantom{${-}$}36.6$\pm$2.3  & <10\textsuperscript{${-}$99}\hphantom{\textsuperscript{\textdagger}}  &  & 95\%{}$\in$(92,98)  & <10\textsuperscript{${-}$95}\hphantom{\textsuperscript{\textdagger}}  &  & 98\%{}$\in$(95,99)  & <10\textsuperscript{${-}$94}\hphantom{\textsuperscript{\textdagger}} \tabularnewline
\midrule
 & Full${\to}$NoAR  & \hphantom{0}${-}$1.2$\pm$2.0  & 0.217\textsuperscript{\textdagger}  &  & 49\%{}$\in$(44,55)  & 0.735\textsuperscript{\textdagger}  &  & 50\%{}$\in$(44,56)  & 0.869\textsuperscript{\textdagger} \tabularnewline
 & Full${\to}$NoPCA  & \hphantom{${-}$}12.0$\pm$1.8  & <10\textsuperscript{${-}$32}\hphantom{\textsuperscript{\textdagger}}  &  & 77\%{}$\in$(72,81)  & <10\textsuperscript{${-}$28}\hphantom{\textsuperscript{\textdagger}}  &  & 71\%{}$\in$(65,77)  & <10\textsuperscript{${-}$10}\hphantom{\textsuperscript{\textdagger}} \tabularnewline
 & Full${\to}$NoFiLM  & \hphantom{${-}$0}2.1$\pm$1.7  & 0.023\hphantom{\textsuperscript{\textdagger}}  &  & 58\%{}$\in$(53,64)  & 0.003\hphantom{\textsuperscript{\textdagger}}  &  & 60\%{}$\in$(53,66)  & 0.003\hphantom{\textsuperscript{\textdagger}} \tabularnewline
Q3 & Full${\to}$NoAudio  & \hphantom{0}${-}$2.2$\pm$1.6  & 0.016\hphantom{\textsuperscript{\textdagger}}  &  & 42\%{}$\in$(36,47)  & 0.003\hphantom{\textsuperscript{\textdagger}}  &  & 33\%{}$\in$(26,39)  & <10\textsuperscript{${-}$6\hphantom{0}}\hphantom{\textsuperscript{\textdagger}} \tabularnewline
 & Full${\to}$NoText  & ${-}$11.2$\pm$1.6  & <10\textsuperscript{${-}$35}\hphantom{\textsuperscript{\textdagger}}  &  & 22\%{}$\in$(17,26)  & <10\textsuperscript{${-}$31}\hphantom{\textsuperscript{\textdagger}}  &  & 11\%{}$\in$(7,15)  & <10\textsuperscript{${-}$44}\hphantom{\textsuperscript{\textdagger}} \tabularnewline
 & Full${\to}$NoVel  & \hphantom{${-}$0}2.6$\pm$1.7  & 0.007\hphantom{\textsuperscript{\textdagger}}  &  & 53\%{}$\in$(47,59)  & 0.459\textsuperscript{\textdagger}  & $-$ & 60\%{}$\in$(53,67)  & 0.006\hphantom{\textsuperscript{\textdagger}} \tabularnewline
 & NoPCA${\to}$GT  & \hphantom{${-}$}37.4$\pm$2.2  & <10\textsuperscript{${-}$99}\hphantom{\textsuperscript{\textdagger}}  &  & 96\%{}$\in$(94,98)  & <10\textsuperscript{${-}$99}\hphantom{\textsuperscript{\textdagger}}  &  & 98\%{}$\in$(96,100)  & <10\textsuperscript{${-}$95}\hphantom{\textsuperscript{\textdagger}} \tabularnewline
\midrule 
 & Full${\to}$NoAR  & \hphantom{0}${-}$4.6$\pm$2.1  & <10\textsuperscript{${-}$4\hphantom{0}}\hphantom{\textsuperscript{\textdagger}}  & $\mathbf{+}$ & 41\%{}$\in$(36,46)  & <10\textsuperscript{${-}$3\hphantom{0}}\hphantom{\textsuperscript{\textdagger}}  & $\mathbf{+}$ & 47\%{}$\in$(41,52)  & 0.197\textsuperscript{\textdagger} \tabularnewline
 & Full${\to}$NoPCA  & \hphantom{${-}$}10.0$\pm$2.0  & <10\textsuperscript{${-}$21}\hphantom{\textsuperscript{\textdagger}}  &  & 71\%{}$\in$(66,76)  & <10\textsuperscript{${-}$17}\hphantom{\textsuperscript{\textdagger}}  &  & 69\%{}$\in$(63,75)  & <10\textsuperscript{${-}$9\hphantom{0}}\hphantom{\textsuperscript{\textdagger}} \tabularnewline
 & Full${\to}$NoFiLM  & \hphantom{${-}$0}2.0$\pm$1.8  & 0.043\hphantom{\textsuperscript{\textdagger}}  &  & 55\%{}$\in$(50,61)  & 0.050\textsuperscript{\textdagger}  & $\mathbf{-}$ & 59\%{}$\in$(52,65)  & 0.008\hphantom{\textsuperscript{\textdagger}} \tabularnewline
Q4 & Full${\to}$NoAudio  & \hphantom{0}${-}$3.8$\pm$1.7  & <10\textsuperscript{${-}$4\hphantom{0}}\hphantom{\textsuperscript{\textdagger}}  &  & 38\%{}$\in$(33,43)  & <10\textsuperscript{${-}$5\hphantom{0}}\hphantom{\textsuperscript{\textdagger}}  &  & 33\%{}$\in$(27,40)  & <10\textsuperscript{${-}$7\hphantom{0}}\hphantom{\textsuperscript{\textdagger}} \tabularnewline
 & Full${\to}$NoText  & ${-}$12.0$\pm$1.6  & <10\textsuperscript{${-}$40}\hphantom{\textsuperscript{\textdagger}}  &  & 20\%{}$\in$(16,25)  & <10\textsuperscript{${-}$34}\hphantom{\textsuperscript{\textdagger}}  &  & 14\%{}$\in$(10,19)  & <10\textsuperscript{${-}$38}\hphantom{\textsuperscript{\textdagger}} \tabularnewline
 & Full${\to}$NoVel  & \hphantom{${-}$0}1.7$\pm$1.7  & 0.043\hphantom{\textsuperscript{\textdagger}}  &  & 51\%{}$\in$(46,57)  & 0.623\textsuperscript{\textdagger}  & $\mathbf{-}$ & 58\%{}$\in$(51,64)  & 0.026\hphantom{\textsuperscript{\textdagger}} \tabularnewline
 & NoPCA${\to}$GT  & \hphantom{${-}$}36.2$\pm$2.4  & <10\textsuperscript{${-}$99}\hphantom{\textsuperscript{\textdagger}}  &  & 94\%{}$\in$(91,97)  & <10\textsuperscript{${-}$89}\hphantom{\textsuperscript{\textdagger}}  &  & 96\%{}$\in$(93,98)  & <10\textsuperscript{${-}$83}\hphantom{\textsuperscript{\textdagger}} \tabularnewline
\bottomrule
\end{tabular}
\end{table*}
\end{document}